\documentclass[prodmode,hillsideplop]{acmlarge}
\usepackage{flafter}
\usepackage{hyperref}
\usepackage[hyphenbreaks]{breakurl}

\usepackage[ruled]{algorithm2e}
\SetAlFnt{\algofont}
\SetAlCapFnt{\algofont}
\SetAlCapNameFnt{\algofont}
\SetAlCapHSkip{0pt}
\IncMargin{-\parindent}
\renewcommand{\algorithmcfname}{ALGORITHM}

\title{A Pattern Sequence for Designing Blockchain-Based Healthcare Information Technology Systems}
\author{Peng Zhang, Ph.D. \affil{Belmont University, Vanderbilt University}\ \\
Douglas C. Schmidt, Ph.D. \affil{Vanderbilt University}\ \\
Jules White, Ph.D. \affil{Vanderbilt University}}

\begin{abstract}

Known for its decentralized and tamper-aware properties, blockchain is attractive to enhance the infrastructure of systems that have been constrained by traditionally centralized and vendor-locked environments. Although blockchain has commonly been used as the operational model behind cryptocurrency, it has far more foreseeable utilities in domains like healthcare, where efficient data flow is highly demanded. Particularly, blockchain and related technologies have been touted as foundational technologies for addressing healthcare interoperability challenges, such as promoting effective communications and securing data exchanges across various healthcare systems. Despite the increasing interests in leveraging blockchain technology to improve healthcare infrastructures, a major gap in literature is the lack of available recommendations for concrete architectural styles and design considerations for creating blockchain-based apps and systems with a healthcare focus. 

This research provides two contributions to bridge the gap in existing research. First, we introduce a pattern sequence for designing blockchain-based healthcare systems focused on secure and at-scale data exchange. Our approach adapts traditional software patterns and proposes novel patterns that take into account both the technical requirements specific to healthcare systems and the implications of these requirements on naive blockchain-based solutions. Second, we provide a pattern-oriented reference architecture using an example application of the pattern sequence for guiding software developers to design interoperable (on the technical level) healthcare IT systems atop blockchain-based infrastructures. The reference architecture focuses on minimizing storage requirements on-chain, preserving the privacy of sensitive information, facilitating scalable communications, and maximizing evolvability of the system. 

\end{abstract}

\category{D.2.7}{Software Engineering}{Distribution and
Maintenance}[Documentation]

\terms{Design, Documentation}
\keywords{Blockchain Technology, Software Engineering, Design Patterns, Smart Contracts, Smart Contract Security and Vulnerability ,  Healthcare, Data Sharing, Interoperability, Solidity}

\acmformat{Zhang, P.\ Schmidt, D.C. \ and White, J. 2020. A Pattern Sequence for Designing Blockchain-Based Healthcare Information Technology Systems.}

\copyr{PLoP'19, October 7--10, Ottawa, Ontario, {Canada}. Copyright 2019 is held by the author(s). {HILLSIDE} 978-1-941652-14-5}

\begin{document}

\begin{bottomstuff}
Author's address: Zhang, Peng\ 1900 Belmont Blvd, Nashville, TN 37212; email: peng.zhang@vanderbilt.edu;
Schmidt, Douglas C.\ Vanderbilt University, PMB 351679, 2301 Vanderbilt Place, Nashville, TN 37235; email: d.schmidt@vanderbilt.edu;
White, Jules \ Vanderbilt University, PMB 351679, 2301 Vanderbilt Place, Nashville, TN 37235; email: jules.white@vanderbilt.edu\\

Permission to make digital or hard copies of all or part of this work for personal or classroom use is granted without fee provided that copies are not made or distributed for profit or commercial advantage and that copies bear this notice and the full citation on the first page. To copy otherwise, to republish, to post on servers or to redistribute to lists, requires prior specific permission. A preliminary version of this paper was presented in a writers' workshop at the 26th Conference on Pattern Languages of Programs (PLoP). PLoP'19, OCTOBER 7-10, Ottawa, Ontario Canada. Copyright 2019 is held by the author(s). HILLSIDE 978-1-941652-14-5
\end{bottomstuff}

\maketitle

\section{Introduction}
Blockchain technology has demonstrated its success in sustaining the secure and scalable exchanges of digital assets through its first applications in cryptographic currency, such as Bitcoin and Ethereum~\cite{nakamoto2008bitcoin,buterin2013ethereum}. In essence, blockchain is a decentralized architecture built upon existing concepts from computer science and mathematics in a manner that differs from traditional infrastructures, which have placed many restrictions on system services and capabilities due to centralization. To achieve decentralization, information becomes transparent and immutable to a certain degree, which enabled the Bitcoin blockchain as a viable platform for "trustless" exchanges~\cite{blundell2014bitcoin} that take place directly between any two parties without the involvement of a trusted middleman or another third party. These revolutionary concepts underlying blockchain have sparked a lot of interest in its application from technologists and domain experts across various industries, such as finance, healthcare, transactive energy, and the food industry.  

Another mainstream public blockchain platform, Ethereum, extended the capabilities of cryptocurrency-based blockchains like Bitcoin to enable programmability and near Turing-complete computations via ``smart contracts``~\cite{buterin2013ethereum}. Smart contracts are similar to any software program in that they have states and some instructions to directly control the states to facilitate the exchange and/or redistribution of digital assets between two or more parties.  The instructions or code define rules or agreements established between involved parties in advance to produce deterministic outputs. Ethereum's successful implementation of programmable smart contracts promoted the development of decentralized apps (DApps)~\cite{johnston2014general}, which are autonomously operated services that interact with cryptography-protected data stored on the blockchain and persist the records of transactions also on-chain. DApps are also a medium that allows end users to directly interact with the blockchain and relevant data on-chain.

Blockchain and smart contracts have been explored as a foundational technologies to address healthcare interoperability challenges~\cite{desalvo2015connecting,das2017does}. Interoperability is the ability for various information technology systems and applications to communicate, exchange data, and effectively digest the exchanged information~\cite{geraci1991ieee}. Healthcare authorities and experts have attempted to improve the interoperability in healthcare for decades~\cite{richesson2011data} in order to provide more continuous and consistent medical services, including but not limited to securely and reliably delivering patient data across different episodes of care and care locations, effectively facilitating medical communications between providers, and accurately and promptly connecting medical devices and medical alerts to the appropriate patients~\cite{lesh2007medical}. However, despite the growing interest in creating blockchain-based healthcare systems, little information exists in current literature on the concrete design recommendations for applying blockchain technology to address healthcare-specific challenges. 

This paper focuses on addressing this research gap by \textit{providing a basic software pattern sequence for designing blockchain-based healthcare DApps that target healthcare-specific challenges}. The target audience of this paper are health information technology (IT) system architects and developers interested in applying blockchain and related technologies in the system design. In software engineering practice, design patterns offer general and reusable solutions to recurring problems. They allow software engineers to communicate using well-known and well-understood names for interactions in the software~\cite{shvets2015design}. By documenting a sequence of reoccurring blockchain-specific patterns taking into account domain-specific requirements, this work can assist the target audience to more quickly adapt to this technology and create robust solutions with it in the healthcare domain.

The remainder of this paper is organized as follows: Section~\ref{sec:concepts} provides an overview of blockchain technology and the Ethereum implementation; Section~\ref{sec:related} summarizes existing research related to this work; Section~\ref{sec:challenges} outlines key challenges regarding healthcare interoperability faced by the direct applications of blockchain technology; Section~\ref{sec:sequence} presents our proposed pattern sequence in detail; 
and Section~\ref{sec:conclusion} concludes the paper and summarizes future work on applying blockchain and related technologies in the healthcare domain.
\section{Key Concepts of Blockchain Technology}
\label{sec:concepts}
This section gives a general overview of blockchain and the open-source Ethereum implementation that supports smart contracts, which enable computation and facilitate the development of decentralized apps beyond cryptocurrencies. The overview is followed by a discussion of Solidity, which is a programming language for writing smart contracts in Ethereum and the basis of our pattern sequence discussed in Section~\ref{sec:sequence}.

\subsection{Blockchain Concepts}
\label{sec:overview}
\begin{figure}[bpth]
\centering
\includegraphics[width=0.75\textwidth]{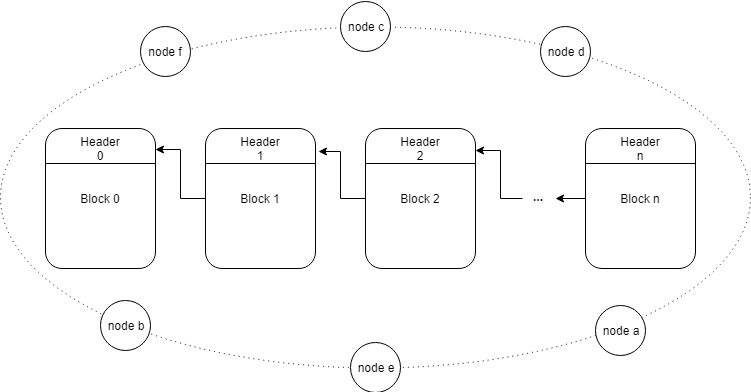}
\caption{Blockchain Structure: A Shared, Append-Only List of Ordered Transactions Grouped into Blocks}
\label{fig:blockchain}
\end{figure}

A blockchain is a decentralized, replicated, and continually reconciled data structure that maintains an append-only list of ordered transactions grouped into blocks, as shown in Figure ~\ref{fig:blockchain}. 

All transactions are recorded in the blockchain and available to all network nodes to provide transparency of the exchanges of digital data, such as cryptocurrencies. Only one block may be added to the blockchain at a time following a mathematical verification (based on cryptography) to ensure that it is in sequence from the previous block and contain all valid transactions. The most famous and robust verification process is \textit{Proof of Work}(PoW) or ``mining``~\cite{nakamoto2008bitcoin} that is introduced in the Bitcoin blockchain. In PoW, block-mining nodes compete to have their block of transactions be the next one added to the blockchain by solving a hard puzzle whose solution is trivial to verify. The first node to solve the puzzle announces the solution to the entire network, and if the solution is correct, the winner receives a mining reward paid via cryptocurrency for their contribution. PoW is based on cryptography, game theory, and incentive engineering to ensure decentralized consensus is reached regarding the block and transaction sequence and to protect the transaction history against tampering. The mining process provides blockchain with three key properties: transparency (for easily verifying that a specific transaction occurred at a particular point in time), immutability (that prevents existing transaction history from being altered), and decentralization (for providing resiliency to a single point of failure with replicated storage)~\cite{cmublockchain}.

In Bitcoin, blockchain serves as a public ledger that facilitates the direct financial transactions between individual users with cryptography to secure the exchange. However, the blockchain implemented by Bitcoin is limited to only support Bitcoin transactions and not suitable for other types of more complex data exchange beyond cryptocurrency. To provide a more flexible framework, Ethereum was created as an alternative blockchain that enables computation, making it a more generalized trustless platform that can run smart contracts~\cite{buterin2013ethereum}. 

Ethereum is a decentralized computing system that has a native cryptocurrency powering the network and an extended programmable capability enabled by the use of the near Turing-complete Ethereum Virtual Machine (EVM). In Ethereum blockchain, a smart contract can be created to store data of different structures and define protocols that interact with the data. To ensure the same level of information consistency and consensus as its Bitcoin predecessor, Ethereum enforces a payment policy in terms of ``gas" for creating and storing each smart contract and any data operation. This policy serves two purposes: (1) it is used to pay for network nodes to verify and execute valid transactions as an incentive and (2) it is also a financial disincentive against malicious attacks.  In addition, there is a global maximum gas limit defined by the Ethereum protocol and a sender-specified gas limit that indicates the max gas amount that the sender is willing to pay for. If gas spent during the execution of a transaction exceeds either of these two limits, computation will be stopped, and the sender still has to pay for the performed computation. This protocol protects senders from completely running out of funds and also further deters malicious attacks and abuse, such as distributed denial of service attacks in the network or hostile infinite loops in smart contract code~\cite{buterin2013ethereum}. 

\subsection{Overview of Solidity}
Ethereum smart contracts can be written in a Turing-complete programming language, called Solidity~\cite{solidity2017ethereum}, compiled by the EVM. Programmable smart contracts foster the development of DApps on the Ethereum blockchain and are thus a feature we are exploring to potentially address certain healthcare interoperability challenges. Solidity is an object-oriented language and is designed primarily for writing contracts in Ethereum. 

A Solidity class is realized through a "contract," which is an object prototype (some code template) stored on-chain. Just like an object-oriented class can be instantiated into a concrete object at runtime, a contract may be instantiated into a concrete ``smart contract account`` (SCA) by a transaction or a function call from another contract. At instantiation, a contract is assigned a unique address that is similar to a reference or pointer in C/C++-like languages. The contract can be referred to and its functions revoked using the address. A smart contract can also define state variables to store data and functions that interact with the data. Although one contract can be instantiated into many SCAs, it should be treated as a singleton to avoid undesired behavior and storage overhead. A common practice is to store the address of an instantiated contract in a static location, such as a configuration file or database. The address can then be retrieved as a parameter to access the contract's internal states and invoke its functions~\cite{dourlens}. 

Solidity also supports multiple inheritance and polymorphism~\cite{soliditypoly}. When a contract inherits from one or more other contracts, a new contract instance is created by copying all the base contracts code into the child contract prototype. An abstract contract in Solidity can declare function headers but without concrete implementations, which means that it cannot be compiled into an SCA but can be used as a base contract. In this paper, we focus the pattern discussions based on the use of Solidity.

\section{Related Work}
\subsection{Overview}
\label{sec:related}
Although relatively few papers focus on realizing software patterns in blockchains, some relate to healthcare blockchain solutions and design principles in this space. This section gives an overview of related research on (1) the challenges of applying blockchain-based technology in the healthcare space and innovative implementations of blockchain-based healthcare systems and (2) design principles and recommended practice for blockchain application implementations.

\subsubsection{Challenges of healthcare blockchain and proposed solutions.}
Azaria et al.~\cite{azaria2016medrec} proposed MedRec as an innovative, working healthcare blockchain implementation for handling EHRs, based on principles of existing blockchains and Ethereum smart contracts. The MedRec system uses database "Gatekeepers" for accessing a node's local database governed by permissions stored on the MedRec blockchain. Peterson et al.~\cite{Peterson2016} presented a healthcare blockchain with a single centralized source of trust for sharing patient data, introducing \textit{Proof of Interoperability} based on conformance to the FHIR protocol as a means to ensure network consensus.

\subsubsection{Prior efforts focused on software design practice for developing blockchain apps.} Porru et al.~\cite{porru2017blockchain} highlighted evident challenges in state-of-the-art blockchain-oriented software development by analyzing open-source software repositories and addressed future directions for developing blockchain-based software. Their work focused on macro-level design principles such as improving collaboration, integrating effective testing, and evaluations of adopting the most appropriate software architecture. Bartoletti et al.~\cite{bartoletti2017empirical} surveyed the usage of smart contracts and identified nine common software patterns shared by the studied contracts, \textit{e.g.}, using "oracles" to interface between contracts and external services and creating "polls" to vote on some question. These patterns summarize the most frequent solutions to handle some repeated scenarios. A number of attacks on Ethereum smart contracts have been reported, including the infamous DAO attack~\cite{siegel2016understanding} where ~\$50 million worth of Ether was stolen and the critical Parity wallet hack~\cite{zepplin2018parity} that incurred in ~\$30 million worth of Ether being exploited.  Atzei et al. surveyed existing attacks on Solidity smart contracts with code snippets showing related vulnerabilities~\cite{atzei2017survey}.  Meanwhile, the blockchain community also compiled a number of software patterns and anti-patterns targeting Solidity programming around cryptocurrency transactions in order to maximize the security of Ethereum smart contract design~\cite{consensys2018practice}.  More recently, Moreno et al. proposed a security pattern focusing on the use of blockchain in big data systems~\cite{moreno2019}. Relatedly, Ellervee et al. described a comprehensive reference model for blockchain-based systems using software architecture concepts~\cite{ellervee2017comprehensive}.

\subsection{Gaps in Existing Research}
Many research and engineering ideas have been proposed to apply blockchain technology in healthcare, and implementation attempts are underway~\cite{azaria2016medrec,Peterson2016,porru2017blockchain,bartoletti2017empirical}. Prior research efforts have provided a number of design recommendations for implementing Solidity smart contracts involving cryptocurrency transactions. Few published studies, however, have addressed software design considerations needed to implement blockchain-based healthcare apps effectively. While it is crucial to understand the fundamental properties of blockchains and the smart contract programming language, it is also important to apply them properly so that healthcare-specific challenges are addressed. Even though a subset of principles from prior work may be relevant to the healthcare space, a systematic approach to document appropriate design practice that specifically target technical challenges in healthcare is still essential.

\section{Healthcare Interoperability Challenges Faced by Blockchain-Based Apps}
\label{sec:challenges}
The US Office of the \textit{National Coordination for Health Information Technology} (ONC) has outlined basic technical requirements for achieving interoperability~\cite{onc2014vision}.  Based on these requirements, this section summarizes key interoperability challenges faced by blockchain-based apps, focusing on four aspects: system evolvability, blockchain storage, information privacy, scalability, and security. 

\subsection{Evolvability Challenge: Maintaining Evolvability While Minimizing Integration Complexity}

Many traditionally centralized apps are written with the assumption that data is easy to change.  This assumption does not hold true for blockchain-based apps.  Once stored on-chain, data is difficult to modify \textit{en masse}.  Not only is code manipulating the data is immutable, but data change history also persists on-chain and can be replayed due to the nature of blockchain.  Healthcare data contains sensitive personal information protected by law~\cite{centers2003hipaa}, which, if compromised, would create severe legal, financial, and also social consequences.  The vulnerability in smart contract code leading to the infamous DAO attack~\cite{siegel2016understanding} must be avoided in a healthcare app.

At the same time, healthcare systems may be subject to updates or upgrades required by clinical workflow or healthcare regulations.  This need for potential system evolution creates a tension for a blockchain-based design.  As such, a critical design consideration when building blockchain apps for healthcare is to ensure that the data written into blockchain via smart contracts are designed to facilitate evolution where mandated.  

Although evolution must be supported, healthcare data must often be accessible from a variety of deployed systems that cannot easily be changed over time.  Apps should therefore be designed in a way that is loosely coupled and minimizes the usability impact of evolution on the clients, \textit{i.e.}, user services that interact with data in the blockchain. Sections~\ref{pattern:ring},~\ref{pattern:manager}, and~\ref{pattern:guarded} shows how using \textsc{Layered Ring}, \textsc{Contract Manager}, and \textsc{Guarded Update} patterns from the pattern sequence, respectively, can help avoid serious attacks like the DAO~\cite{siegel2016understanding} and facilitate necessary system evolution, while minimizing the impact on dependent clients, focusing on the separation of concerns between data and logic and a type of attack called reentrancy, which will be described further in Section~\ref{pattern:guarded}.
 
\subsection{On-Chain Storage Challenge: Minimizing Data Storage Requirements on the Blockchain}
Healthcare apps can serve thousands to millions of participants, which may incur enormous overhead when \textit{large} volumes of data are stored in a blockchain--particularly if data normalization and denormalization techniques are not carefully considered. Considering storage scalability, not only is it costly to store data, but data modifications and access operations may also fail if/when the cost of storage or execution exceeds the allowance in a blockchain, \textit{e.g.}, gas limit defined for the Ethereum blockchain as discussed in Section~\ref{sec:overview}. An important design consideration for blockchain-based healthcare apps is thus to minimize data storage requirements in addition to provide sufficient flexibility to manage individual health concerns. Sections~\ref{pattern:connector} and \ref{pattern:registry} show how to design smart contracts with \textsc{Database Connector} and \textsc{Entity Registry} patterns from the pattern sequence, respectively, to improve interoperability by standardizing interfaces to storage access and maximizes on-chain scalability by capturing common intrinsic data sharing across entities while still allowing extrinsic data to vary in specific entity contracts.

\subsection{On-Chain Privacy Challenge: Balancing Data Storage with Privacy Concerns}
Blockchains and smart contracts can offer trustless digital health asset sharing, audit trails of data access, and decentralized and replicated storage, which are essential for improving healthcare interoperability by providing ubiquitous data store. 
Although there are substantial potential benefits to the availability of information if data is stored on-chain, there are also significant risks due to the transparency of blockchain. In particular, even when encryption is applied to sensitive data on-chain, it is still possible that the current encryption techniques may be broken in the future~\cite{rich2014nsa} or that vulnerabilities in the encryption implementations may later be exploited, rendering private information potentially decryptable in the future. To protect health information privacy, in Sections~\ref{pattern:proxy} and \ref{pattern:token} we discuss how designing a blockchain-based app using the \textsc{Database Proxy} and \textsc{Tokenized Exchange} patterns from the pattern sequence, respectively, can facilitate data sharing while keeping sensitive patient data from being directly encoded in the blockchain. 
 
\subsection{Scalable Communication Challenge: Tracking Relevant Health Changes Scalably Across Large Patient Populations}

Communication gaps and information sharing challenges are serious impediments to healthcare innovation and the quality of patient care. Providers, hospitals, insurance companies, and departments within health organizations experience disconnectedness caused by delayed or lack of information flow. Patients are commonly cared for by various sources, such as private clinics, regional urgent care centers, and enterprise hospitals. A provider may have hundreds or more patients whose associated health data must be tracked. Section~\ref{pattern:pubsub} shows how a blockchain-based app design using the \textsc{Publisher-Subscriber} pattern from our pattern sequence can be aid in scalably detecting and communicating relevant health changes.

\subsection{Security Challenge: Preventing Unintended Software Loopholes and Safeguarding On-Chain Data}
Similar to information privacy that protects the identity of users and data holders, information and system security is also paramount in healthcare systems to safeguard against unintended software loopholes or unauthorized access that could lead to compromised data. Due to the decentralized nature of blockchain and related technology, the security of information faces higher risks because it is exposed to a much wider, and sometimes uncontrolled, audience and it is usually not immediately amendable.  In recent years, software loopholes that existed in several major blockchain-based crytocurrency services have been exploited by attackers, causing significant financial losses to users and the service providers~\cite{}.  To prevent similar predicament from happening to healthcare services hosted in blockchain-based infrastructures, security risks must be recognized in the early stage of system designs in addition to the construction of a safeguarded system.  Sections~\ref{pattern:guarded} and~\ref{pattern:pubsub} are two examples of security protection in a blockchain-based healthcare app. 

\section{A Key Pattern Sequence for Designing Blockchain-Based Health Apps}
\label{sec:sequence}
This section presents a key pattern sequence for creating blockchain-based health system designs that address the major challenges described earlier in Section~\ref{sec:challenges}.  Our research approach that developed this sequence has three folds.  First, because the topic of design patterns focused on using blockchain technology for healthcare has received limited attention in literature, we had extracted a subset of the patterns using commonality and variability analysis~\cite{coplien1998commonality}.  Specifically, we obtained a number of verified smart contract source code from Etherscan.io~\cite{ethscan} to capture common portions repeatedly used across various contracts and/or supporting library contracts, which we codified into patterns of this sequence, such as \textsc{Layered Ring} and \textsc{Contract Manager}  Second, based on our experience from previous work on researching healthcare data sharing solutions~\cite{zhang2017applying,ZHANG2018267} and our understanding of the healthcare domain and the technical requirements for its systems~\cite{zhang2017metric,ZHANG2018}, we codified the design practice we learned from prior research into several patterns in the key sequence, such as \textsc{Database Connector} and \textsc{Tokenized Exchange}.  Third, given the extensiveness and maturity of existing research on centralized and distributed software engineering design practice, we have applied, wherever necessary for a blockchain-based healthcare system, design principles widely accepted to our pattern sequence with blockchain-focused design considerations, such as \textsc{Database Proxy, Entity Registry, and Publisher-Subscriber}.  Additionally, due to the growing popularity of Solidity (which is the primary programming language for creating smart contracts) and attacks that have occurred to public smart contracts, the Ethereum community has captured a number of Solidity code patterns for preventing similar attacks.  Although those code patterns were almost exclusively targeting cryptocurrency or other apps with financial incentives, we identified one code pattern that would be particularly critical in a healthcare system, namely, \textsc{Guarded update}.

The remainder of this section applies a pattern form variant to motivate and show how our pattern sequence aids in designing blockchain-based healthcare apps.  In particular, we present eight software patterns---\textsc{Layered Ring, Guarded Update, Contract Manager, Database Connector, Database Proxy, Entity Registry, Tokenized Exchange}, and \textsc{Publisher-Subscriber}~\cite{gamma1995design,buschmann2007pattern}.  We describe key healthcare challenges that they resolve in the blockchain platform and detail their structure and composition. \footnote{Naturally, there are other patterns relevant in this domain, which can be the focus of future work.}.  

Table~\ref{tbl:sequence} provides an overview of the pattern sequence, showing how the patterns relate to healthcare-specific challenges described previously in Section~\ref{sec:challenges} and what specific sub-challenge each pattern aims to solve.

\begin{table}[ht]

\label{tbl:sequence}
\tbl{Overview of Proposed Pattern Sequence for Designing Blockchain-Based Healthcare Apps}{
\begin{tabular}{|l|l|l|}
\hline
Pattern              & Targeted Category                                                 & Specific Challenge to Solve                                                                                                                                             \\ \hline
Layered Ring         & Evolvability                                                      & \begin{tabular}[c]{@{}l@{}}Defining the data sharing system’s base \\ architecture\end{tabular}                                                                         \\ \hline
Guarded Update       & Evolvability \& Security                                                     & \begin{tabular}[c]{@{}l@{}}Preventing unexpected reentrancy attacks \\ that occurred in the DAO\end{tabular}                                                            \\ \hline
Contract Manager     & Evolvability                                                      & \begin{tabular}[c]{@{}l@{}}Separating data from logic to ensure data \\ availability via clean separation of concerns\end{tabular}                                      \\ \hline
Database Connector   & On-Chain Storage                                                  & \begin{tabular}[c]{@{}l@{}}Ensuring on-chain storage scalability and \\ interoperability via standardized and minimal \\ interfaces to off-chain storage\end{tabular}   \\ \hline
Database Proxy       & On-Chain Privacy                                                  & \begin{tabular}[c]{@{}l@{}}Providing an additional layer of security by \\ performing lightweight tasks before permitting \\ access to database connectors\end{tabular} \\ \hline
Entity Registry      & On-Chain Storage                                                  & \begin{tabular}[c]{@{}l@{}}Managing healthcare entities on-chain and other\\ types of common data at scale\end{tabular}                                                 \\ \hline
Tokenized Exchange   & On-Chain Privacy                                                  & \begin{tabular}[c]{@{}l@{}}Authorizing access to data storage and \\ maintaining verifiable access logs\end{tabular}                                                    \\ \hline
Publisher-Subscriber & \begin{tabular}[c]{@{}l@{}}Scalable Communication \\ \& Security \end{tabular} & \begin{tabular}[c]{@{}l@{}}Providing user notifications when events of\\ interest occur across the decentralized network\end{tabular}                                   \\ \hline
\end{tabular}
}
\end{table}

Each of the patterns in this sequence is discussed in depth below.

\subsection{A Blockchain-Based Architecture for Health Data Sharing Systems}
\label{pattern:ring}

\textbf{Design problem faced by blockchain-based apps.}  Healthcare data exists in siloed data warehouses across different healthcare organizations, private practices, and, more recently, mobile health app providers~\cite{ajami2013barriers,ZHANG2018267}.  Despite the adoption of certified EHRs or other data exchange solutions that can provide direct data exchange between providers within the same network (\textit{e.g.}, using an EHR system provided by the same vendor), impediments for healthcare providers and researchers to access those heterogeneous data silos still exist.  

\textbf{Solution $\rightarrow$ Apply the \textsc{Layered Ring} pattern to define the base architecture of the health data sharing system.} 
The emerging blockchain technology that supports decentralized data storage and executable code via smart contracts, with Ethereum~\cite{buterin2013ethereum} being the most popular, has presented itself as a potential infrastructure to connect existing healthcare data silos~\cite{Nichol2016,Broderson2016,Dubovitskaya2017} with its success in maintaining tamper-proof cryptocurrency transactions between worldwide Internet users ~\cite{nakamoto2008bitcoin,cap2018cryptocurrency} and managing verifiable collectibles or rewards from cryptogaming like CryptoKitties~\cite{cryptokitties}.

\begin{figure}[tphb]
\begin{center}
\centerline{\includegraphics[width=0.9\columnwidth]{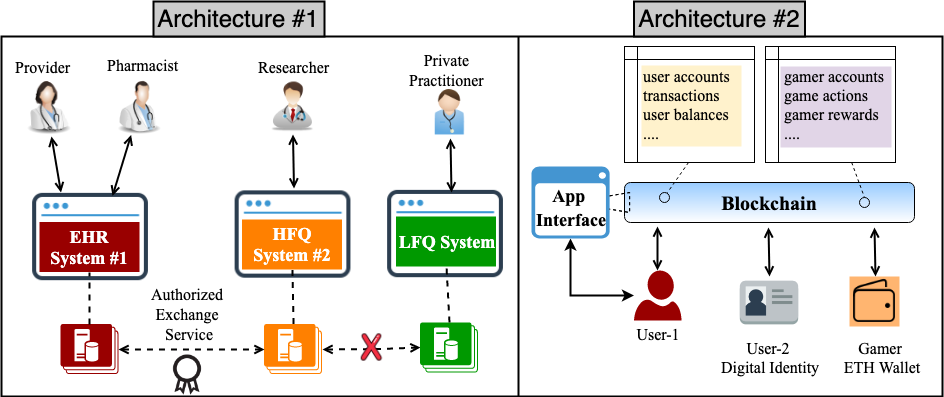}}
\caption{Comparing the Current State of Traditionally Centralized Healthcare Architecture with that of Popular Blockchain-Based Use Cases}
\label{fig:dappcomp}
\end{center}
\end{figure}

At the architectural level, the healthcare data sharing problem is not too different from those successful use cases of blockchain.  Figure~\ref{fig:dappcomp} compares the high-level architecture of data sharing in healthcare with that of blockchain-based cryptocurrency exchange and cryptogaming.  In this figure, the bottom level in both architectures 1 and 2 contains a number of heterogeneously represented objects, i.e., siloed healthcare data sources (e.g., low-frequency, high-fidelity clinical data, or LFQ, captured by trusted sources and high-frequency, low-fidelity data, or HFQ, generated by patients or wearable and mobile devices) in \textit{Architecture 1} and geographically dispersed Internet users in \textit{Architecture 2}.  Data sources generated by healthcare professionals via from diverse, centralized EHR systems on the left may or may not inter-operate, depending on if an authorized exchange service is available between the data sources.  Whereas on the right, data (like identifiers of users/gamers) and data requests flow into and out of the same service implemented on the blockchain that is decentralized and widely accessible, with or without a user interface.  

Consequently, a blockchain-based healthcare data sharing can apply the same basic pattern of \textit{Architecture 2} in Figure~\ref{fig:dappcomp}, except that a user interface is needed for normal healthcare users who typically do not have advanced knowledge about how to execute smart contract functions.  In fact, most blockchain apps, such as CryptoKitties (a cryptogame for collecting and breeding digial cats)~\cite{cryptokitties}, Fomo3D (a gambling game for winning cryptocurrency lotteries)~\cite{fomo3d}, and IDEX (a cryptocurrency trading platform)~\cite{idex}, implement a user-friendly interface that encapsulates the blockchain component, providing users with familiar experience as if interacting with any other centralized web app.  

\begin{figure}[tphb]
\begin{center}
\centerline{\includegraphics[width=0.5\columnwidth]{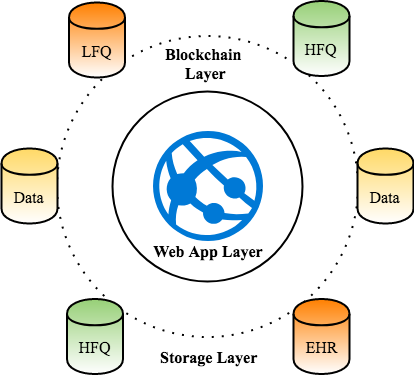}}
\caption{Structure of the \textsc{Layered Ring} Pattern that Defines the Base Architecture of the Data Sharing System}
\label{fig:ring}
\end{center}
\end{figure}

In Figure~\ref{fig:ring} we present the first pattern in the sequence, \textsc{Layered Ring}, generalized from \textit{Architecture 2} above with a bird's eye view to better illustrate the scale of involved entities in each layer.  The outermost layer is a \textit{Storage Layer}, which contains a large number of data sources, each maintained by its owner (e.g., a private practitioner, a healthcare organization, or a 3rd party HFQ data provider).  The middle \textit{Blockchain Layer} connects data sources from the outer layer and would be maintained by key stakeholders or mid- to large-size healthcare organizations in a consortium environment.  The innermost \textit{Web App Layer} provides a convenient interface for interacting with data and operations defined in the blockchain.  It is also the most centralized piece of the system because a web app is usually hosted in a centralized server.  Nevertheless, with a careful design of the web app server, this layer would not introduce additional dependency to the rest of the system and thus maximize the separation of concerns across the overall system.

The \textsc{Layered Ring} pattern is also a variant of the \textsc{Enterprise Service Bus} (ESB) pattern~\cite{zdun2006survey,fernandez2013security}, which provides a common data model and a messaging infrastructure to allow different systems to communicate through a shared set of interfaces. In \textsc{Layered Ring}, the \textit{Blockchain} layer acts as the messaging bus that provides services to the rest of the components. However, unlike the ESB, the interface of \textit{Layered Ring} focuses only on the structural and syntactic level of the exchanged data and does not itself create a common ontology for interpreting the semantics of shared data, which is an entirely separate and complex topic being researched on by domain experts.

After defining the base-line architecture for a blockchain-based data sharing system, we will present other patterns in the sequence applied in each layer to address healthcare-specific challenges described previously in Section~\ref{sec:challenges}.  

\subsection{Preventing Reentrancy Attack in the Blockchain}
\label{pattern:guarded}

\textbf{Design problem faced by blockchain-based apps.}  Despite the growing interest in using blockchain technology for healthcare, a lot of recent attacks on some of the major blockchain-based apps have raised security concerns regarding the use of this technology in especially the healthcare industry that requires compliance to strict security and privacy regulations.  An infamous example of such attacks is the DAO attack~\cite{siegel2016understanding} in which a reentrancy bug was discovered and exploited that causes then worth \$30 million of Ethereum being stolen.  Even though the immutability and decentralization properties of blockchain technology can provide tremendous value to the direct exchange of digital information, without proper design decisions made prior to deploying a system on-chain could yield destructive consequences.

\textbf{Solution $\rightarrow$ Apply the \textsc{Guarded Updated} pattern to prevent unexpected reentrancy attacks.}  We deem attack prevention as the utmost important design consideration in the development cycle of a blockchain-based healthcare system and therefore introduce \textsc{Guarded Updated} as the second pattern in our pattern sequence after defining a base layer with the \textsc{Layered Ring}.  The goal of this pattern is to provide software engineers with a pattern that prevents an important yet serious reentrancy attack early on during the development cycle in order to help design the rest of the system wherever this pattern may apply.  

\begin{figure}[tphb]
\begin{center}
\centerline{\includegraphics[width=0.9\columnwidth]{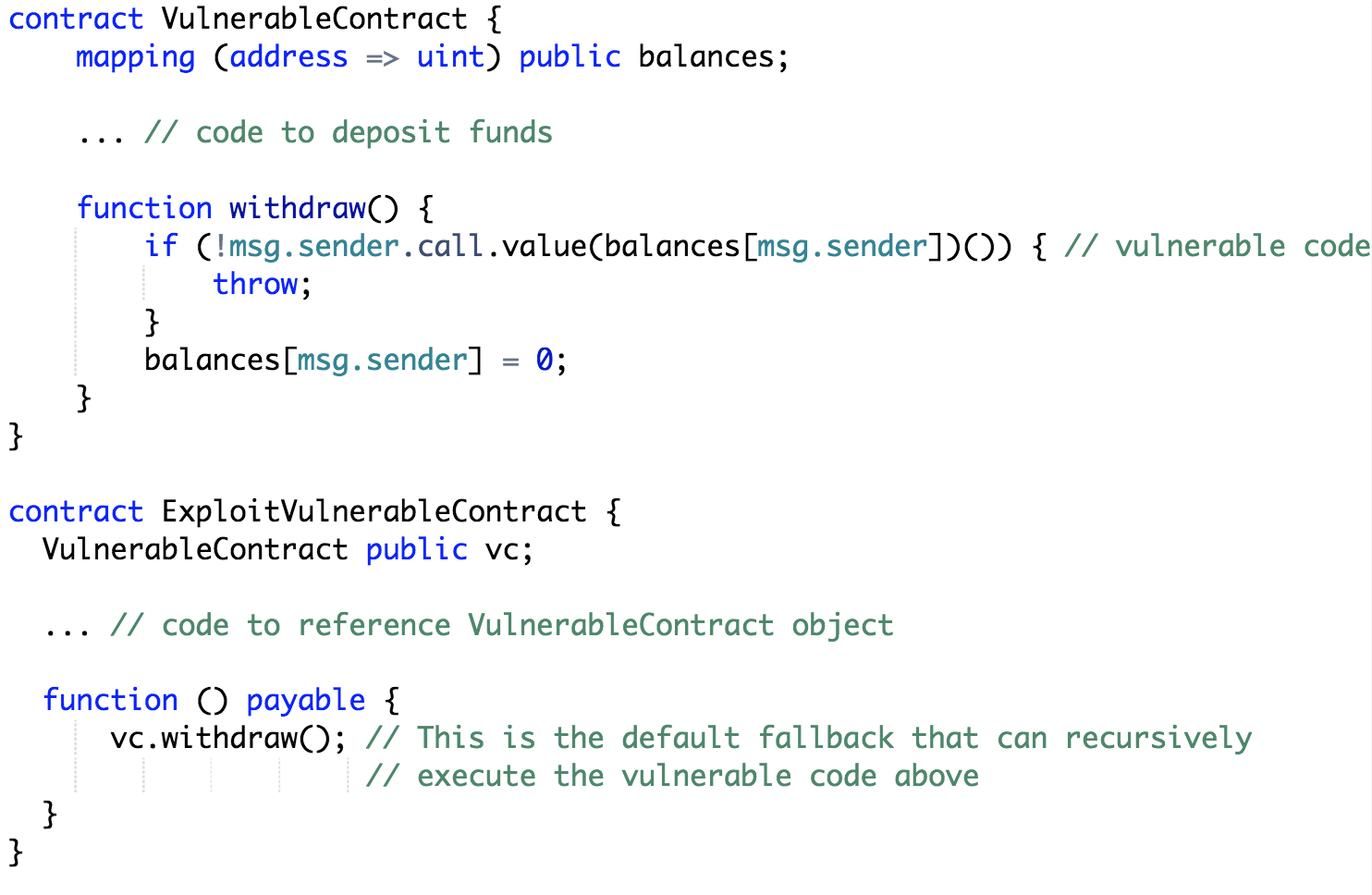}}
\caption{Example Vulnerable Solidity Code of the Simplified Reentrancy Bug and its Exploitation}
\label{fig:code}
\end{center}
\end{figure}

A simplified reentrancy bug that affected the DAO app appears in the code snippet shown in Figure~\ref{fig:code}.  Function \textit{withdraw} in the \textit{VulnerableContract} sets the caller's balance after checking if the asset transfer to the caller (\textit{msg.sender}) is successful.  The attack in \textit{ExploitVulnerableContract} exploits this vulnerability by calling the \textit{withdraw} function in a fallback function that is executed by the \textit{call.value} method, creating recursions that bypass the statement on line that sets the user balance supposedly after the vulnerable statement returns~\cite{consensys2018practice}.

\begin{figure}[tphb]
\begin{center}
\centerline{\includegraphics[width=0.7\columnwidth]{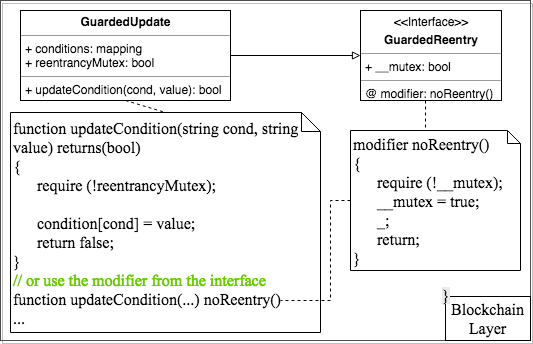}}
\caption{Structure and Example Solidity Code Snippet of \textsc{Guarded Update} Pattern to Prevent Reentrancy Attacks on-Chain}
\label{fig:guarded}
\end{center}
\end{figure}

Although the reentrancy bug primarily targets cryptocurrencies in the interest of gaining financial returns, this bug could also plague systems designs for healthcare functions if prevention is not implemented in advance.  As a key pattern in the sequence, \textsc{Guarded Update} aims to prevent reentrancy attack by ensuring atomic update to critical data in the blockchain-based healthcare system.  The structure and code examples of this pattern appears in Figure~\ref{fig:guarded}\footnote{The code examples are based on https://github.com/o0ragman0o/ReentryProtected}.  As shown in the figure, a boolean guarding condition (i.e., reentrancyMutex) is used to control operations on protected state variable(s) (i.e., \textit{conditions}).  Once the variable(s) has been modified, the guarding condition can be reset to the initial state to permit other memory contexts to act upon the guarded data.  Another, more systematic way to achieve this is to create a modifier in a Solidity interface contract, which can then be included in the declaration header of functions in other contracts.

Protecting atomic updates to state variables in the smart contracts prevents serious reentrancy attacks to occur, however, one major drawback is that atomic executions may slow down runtime performance of the system, particularly in a decentralized environment.

\subsection{Separating Data from Logic to Ensure Data Availability via a Manager Contract}
\label{pattern:manager}

\textbf{Design problem faced by blockchain-based apps.}  The immutability property of blockchains can ensure non-repudiation of data operations and/or transactions of data but can also become a major hurdle to data flow.  On the one hand, immutability is important for achieving interoperability in a healthcare environment as it makes data objects (whether it is a reference pointer to a data store or an authorization request that grants a provider access to healthcare data) on the blockchain always available, even when one of the key maintainers of the network becomes unavailable.  On the other hand, without a loosely-coupled design that focuses on clean separation of data and logic, immutability makes any upgrade to a blockchain-based health system hard to perform.  Data, in such a system, does not only include information being exchanged across various network participants but also needs to contain meta data regarding the system that provides users with the most up-to-date knowledge regarding the system; whereas, logic refers to any operation or event that acts upon the data, typically implemented to read, update, or remove a data object.  

\begin{figure}[th]
\begin{center}
\centerline{\includegraphics[width=0.7\columnwidth]{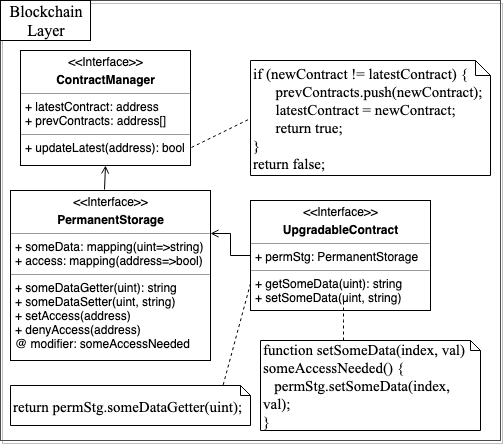}}
\caption{Structure and Example Solidity Code of \textsc{Contract Manager} Pattern for Maintaining Key Meta-Data on-Chain}
\label{fig:manager}
\end{center}
\end{figure}

\textbf{Solution $\rightarrow$ Apply the \textsc{Contract Manager} to separate data from logic to ensure data availability via clean separation of concerns
.}  The \textsc{Contract Manager} pattern aims to address the separation of data and logic via a \textit{permanent storage} structure, which has been described in~\cite{consensys2018practice}.  Figure~\ref{fig:manager} presents the composition of this pattern. 

\textit{Permanent storage} maintains one or more data fields used throughout the system and provides permanent access to data with getter and setter functions for each one of the data fields.  This ensures that all meta data accessed by the system (such as the version or address information of any smart contract dependencies and other data structures shared across different smart contracts) remains readable even when logic contracts are outdated.  Additionally, contract manager stores a \textit{Contract Repository} of meta data that describes versions of the system (including but is not limited to addresses of the latest logic contract components and history contract addresses).  To better ensure upgradeability of the system, \textsc{Contract Manager} also defines access privilege of smart contracts by allowing the original owner of the storage contract to configure an access group for delegating or revoking certain or all rights of accessing or manipulating the data to other members to prevent data locking.

One disadvantage to the introduction of \textsc{Contract Manager} is that all other logic contracts must execute additional calls to this contract for versioning checks and data queries. An alternative design is to leverage the \textsc{Authorizer} pattern~\cite{fernandez2013security} along with fine-grained authorization models such as role-based access control models~\cite{sandhu1996role} or access matrix~\cite{sandhu1994access} to separate the definition of access rules, further decoupling the rules from the storage component. 

\subsection{Standardized On-Chain Interfaces to Off-Chain Storage Access}
\label{pattern:connector}

\textbf{Design problem faced by blockchain-based apps.}  EHR systems have served the U.S. healthcare for decades and, unavoidably, have accumulated enormous amounts of valuable medical records that either exist in legacy systems or in more modern certified EHRs.  Health data sharing today is only possible between healthcare professionals using the same EHR systems or compatible health information exchange services, which are exactly the third-party reliance that blockchain technology helps eliminate with its decentralized, trustless infrastructure.  The direct exchange of digital information on the blockchain is only possible if such information or its representation is encoded on the blockchain with some degree of verifiable integrity.  Due to the scale and privacy of healthcare data, it is unrealistic to store encrypted or hashed version of the actual data on the blockchain.  Furthermore, it is impractical to create a blockchain-based system that completely replaces existing EHR systems or duplicates their functionality.  The design of a scalable and standardized component that connects existing EHR data to a decentralized system offering interoperable data sharing is therefore needed.

\begin{figure}[tphb]
\begin{center}
\centerline{\includegraphics[width=0.7\columnwidth]{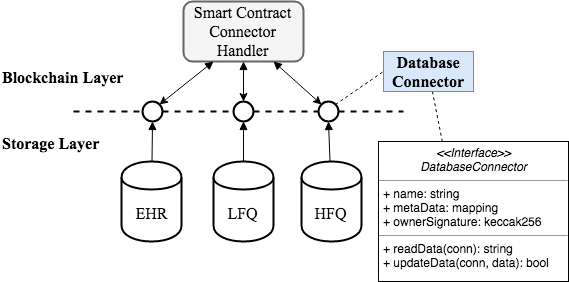}}
\caption{Structure of the \textsc{Database Connector} Pattern Used to Standardize on-Chain Interfaces to off-Chain Storage Access}
\label{fig:connector}
\end{center}
\end{figure}

\textbf{Solution $\rightarrow$ Apply the \textsc{Database Connector} pattern to ensure on-chain storage scalability and interoperability via standardized and minimal interfaces to off-chain storage.}  Figure~\ref{fig:connector} presents the composition of the \textsc{Database Connector} pattern.  The \textit{Database Connector} component defines a standardized interface between the blockchain and storage layers.  The interface provides an abstraction of the heterogeneous health data silos (e.g., EHR or other LFQ databases and HFQ data) to expose only minimal amount of information regarding each data source to the blockchain layer.  As shown in Figure~\ref{fig:connector}, the interface may only need to capture the name or description of a data source, some "meta data" providing reference pointers to the data source, and a verifiable digital signature of the data source owner that provides some level of integrity.  \textit{Database Connector} is also closely associated with the \textsc{Database Proxy} pattern (discussed next in Section~\ref{pattern:proxy}) that uses a \textit{Connector Handler} component in the blockchain layer to provide data access to the connector.

The main benefits of \textsc{Database Connector} are (1) the storage scalability it provides on the blockchain that allows efficient sharing of connectors and (2) a standardized interface that unifies the on-chain representation of off-chain databases.  The drawback is the additional implementations that are required for creating connectors to existing databases.

\subsection{Security Checking before Accessing Off-Chain Storage}
\label{pattern:proxy}

\textbf{Design problem faced by blockchain-based apps.} If a blockchain-based healthcare app must expose sensitive data or metadata (such as patient identifying information) on the blockchain, it must be designed to maximize health data privacy while facilitating health information exchange. In particular, a fundamental aspect of a blockchain is that data and all change history stored on-chain are public, immutable, and verifiable. For financial transactions focused on proving that transfer of an asset occurred, these properties are critical. When the goal is to store data in the blockchain, however, it is important to understand how these properties will impact the use case.

For example, storing patient data in the blockchain can be problematic since it requires that data be public and immutable. Although data can be encrypted before being stored, should all patient data be publicly distributed to all blockchain nodes? Even if encryption is used, the encryption technique may be broken in the future or defects in the implementation of the encryption algorithms or protocols used may make the data decryptable in the future. Immutability, on the other hand, prevents owners of the data from removing the data change history from the blockchain if a security flaw is found. Many other scenarios, ranging from discovery of medical mistakes in the data to changing data standards may necessitate the need to change the data over time.

In scenarios where the data may need to be changed, the public and immutable nature of the blockchain creates a fundamental tension that must be resolved. On the one hand, healthcare providers would like incorruptible data so its integrity is preserved. At the same time, providers want the data changeable and secure to protect patient privacy and account for possible errors. An interoperable app should protect patient privacy and also ensure data integrity.

\textbf{Solution $\rightarrow$ Apply the \textsc{Database Proxy} pattern to provide an additional layer of security by performing lightweight tasks before permitting access to database connectors.}
\textsc{Database Proxy} is akin to the traditional \textsc{Proxy} pattern~\cite{gamma1995design} with a slightly different focus unique to a blockchain-based design.  To reduce computational costs on-chain, the \textit{Database Proxy} interface defines some lightweight representation or placeholder for the real data object and encodes some lightweight security checks or auditing tasks until retrieval of the original data object is required. It is worth noting that because protected health information is only unpackaged or decrypted off-chain, any regulatory or security checking (which is more rigorous), such as authentication or authorization requirements defined by HIPAA or other privacy standards, is performed off-chain instead of on-chain.

\begin{figure}[tphb]
\begin{center}
\centerline{\includegraphics[width=0.7\columnwidth]{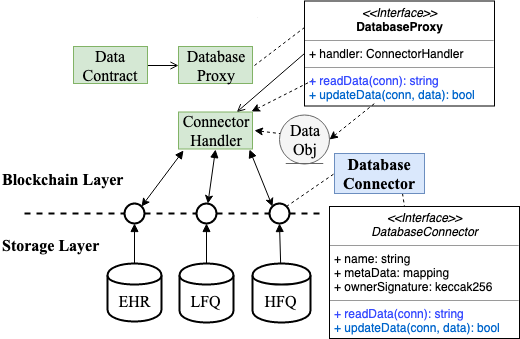}}
\caption{Composition of \textsc{Database Proxy} Pattern for Performing Additional Security Checks before Accessing off-Chain Data Store}
\label{fig:proxy}
\end{center}
\end{figure}

 
Figure~\ref{fig:proxy} illustrates the structure of \textsc{Database Proxy} pattern and its interaction with the \textit{Database Connector} object described previously in Section~\ref{pattern:connector}.

The \textit{Database Proxy} interface maintains a reference to a \textit{Connector Handler} object that forwards the read and write access to the appropriate \textit{Database Connector} for access databases in the storage layer of the system.  Each read request and modification operation through the \textit{Connector Handler} can be logged in an immutable audit trail that is transparent to the entire blockchain network for verification against data corruption.  In the case of a proxified contract (i.e., \textit{Database Connector} that has somewhat heavyweight implementation) being updated with a new storage configuration (\textit{e.g.}, when a data source has been introduced a new management system that requires some change in its \textit{Database Connector} abstraction), the interface to the proxy contract can remain unchanged, encapsulating lower-level implementation variations. 

As with the traditional \textsc{Proxy} pattern~\cite{gamma1995design}, a proxy object can perform lightweight housekeeping operations, such as security checks of administrative access and auditing tasks that log existing data requests, by storing some commonly used metadata in its internal states before  retrieving the actual data.  This component follows the same interface as the real object and can execute the original data object's function implementations as needed.  It provides an additional layer for securing access to the real data object.  However, \textit{Database Proxy} may cause disparate behavior when the real object is accessed directly by some other component in the system while the proxy surrogate is accessed by others.  It also creates an additional level of indirection for accessing actual data objects.

\subsection{Managing Healthcare Entities and Other Types of Common Data on-Chain at Scale}
\label{pattern:registry}

\textbf{Design problem faced by blockchain-based apps.}  All data and transaction records maintained in the blockchain are replicated and distributed to every node in the network.  In a public blockchain, to compensate blockchain miners for contributing expensive hardware to store and maintain on-chain data, fees are charged based on the storage requirement of an application.  Although a fee is not necessarily charged in a consortium blockchain with like-minded parties, other forms of compensation may exist to provide some incentives for the decentralized network maintainers.  To minimize on-chain storage burden, a blockchain-based healthcare app that requires storage of some data on-chain must maximize data sharing among entities thus limit the amount of information stored.

In a large-scale healthcare setting, if a blockchain is used to store patient billing data, there will be millions of records replicated on all blockchain miner nodes. Moreover, billing data could include detailed patient insurance information, such as their ID\#, insurance contact information, coverage details, and other aspects that the provider needs to bill for services. Capturing all this information for every patient can generate excessive amounts of data in the blockchain. 

Suppose it is necessary to store a patient's insurance and billing information (encrypted) in the blockchain. Most patients are covered by one of a relatively small subset of insurers (in comparison to the total number of patients, \textit{e.g.}, each insurance policy may cover 10,000s or 100,000s of patients). Therefore, a substantial amount of intrinsic, non-varying information is common across patients that can be reused and shared, such as details on what procedures are covered by an insurance policy. To bill for a service, however, this common intrinsic information must be combined with extrinsic information (such as the patient's ID\#) that is specific to each patient. 

A good design should maximize sharing of such common data to reduce on-chain storage cost and meanwhile have the capability to provide complete data objects on demand.

\begin{figure}[tphb]
\begin{center}
\centerline{\includegraphics[width=0.7\columnwidth]{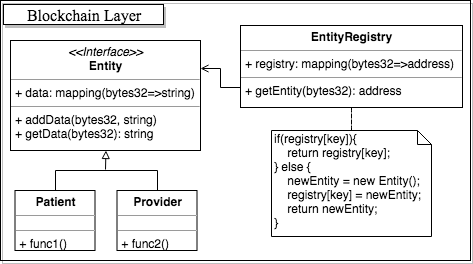}}
\caption{\textsc{Entity Registry} Pattern Used with a Factory to Manage Entities and Other Types of Common Data while Minimizing On-Chain Storage Requirements}
\label{fig:registry}
\end{center}
\end{figure}

\textbf{Solution $\rightarrow$ Apply the \textsc{Entity Registry} pattern for managing healthcare entities on-chain at scale.} 
As shown in Figure~\ref{fig:registry}, the \textsc{Entity Registry} mimics the traditional \textsc{Flyweight} pattern~\cite{gamma1995design} with a factory~\cite{gamma1993design} object to help manage healthcare entities on-chain at scale.  In particular, \textit{getEntity} uses a factory to create entity objects and maintain references (addresses) to created Entity objects in a common smart contract (i.e., \textit{Entity Registry}).  It internalizes common data across a number of \textit{Entity}'s \textit{data} field  while externalizing varying data storage in entity-specific contracts (such as \textit{Patient} or \textit{Provider} entity).  Using references (\textit{i.e.}, addresses) to entity-specific contracts stored in the \textit{registry}, combined extrinsic and intrinsic data can be retrieved upon request to return a complete dataset.

Applying this pattern to the earlier scenario, shared patient insurance information is stored only once in the \textit{registry}, avoiding an exorbitant amount of memory usage from saving repeated data in all patient accounts. Varying, patient-specific billing information is stored in corresponding patient-specific entity contracts. 

The registry can also maintain a mapping between unique entity identifiers and the referencing addresses of already deployed entity contracts to prevent account duplication. At account creation, only if no account with the specified entity identifier exists in the registry does it deploy a new entity contract; otherwise the registry retrieves the address associated with the existing entity contract. To retrieve complete insurance and billing information of a particular patient, clients need only invoke a function call from the registry with the patient identifier to obtain the combined intrinsic and extrinsic data object.

\textsc{Entity Registry} provides better management for the large pool of objects (such as user accounts in the example above). It minimizes redundancy in similar objects by maximizing data and operation sharing. Particularly in the insurance example, if common insurance policy details are not extracted from each patient's contract, the cost to change a policy detail will be immense--it will require rewriting a huge number of impacted contracts. Data sharing with flyweight registry helps minimize the cost to change the common state in objects stored on-chain.

Although applying the \textsc{Entity Registry} pattern creates an additional transaction to verify and include in the blockchain (\textit{i.e.}, the flyweight object instantiation) before it can be used, this extra step can be outweighed by the resulted efficiency in data management. The application of this pattern alone does not ensure integrity of the data being exchanged because it exposes only reference information for retrieving actual data objects for security and privacy reasons. It would rely on an off-chain or a 3rd party oracle service~\cite{xu2016blockchain} to certify the integrity of the data either via hashing functions or other data verification protocols.

\subsection{Securing and Recording Data Access}
\label{pattern:token}

\textbf{Design problem faced by blockchain-based apps.}  Smart contracts are powerful for automating executions of predefined agreements directly between involved entities especially when the entities are registered on the blockchain using its native cryptographic keys and agreed terms are simple updates to cryptocurrency wallets/balances that are easy to update.  The direct exchange of healthcare data unfortunately cannot easily be achieved on-chain due to the complexity and variability in the warehouses and management systems data resides in.  Even when data sharing is made possible in such a decentralized environment, the shared information should not be available to the entire network, unlike an app that involves cryptocurrency.  Instead, proper authorizations of sensitive health data access must be safeguarded.

\textbf{Solution $\rightarrow$ Apply the \textsc{Tokenized Exchange} pattern to authorize access to off-chain data storage and maintain a verifiable data access history.}  Variability of off-chain data sources can be encapsulated with a standardized interface that encodes a set of attributes describing the sources and some basic operations acting upon them (i.e., functions to retrieve the original data source and verify the digital signature to ensure data is originated from the expected sender.).  Figure~\ref{pattern:token} presents the structure of the \textsc{Tokenized Exchange} pattern in the sequence that defines a \textit{Token} interface off-chain to represent each data source in a more consistent manner.  With this interface, the \textit{Database Connector Object} from the \textsc{Database Connector} pattern discussed in Section~\ref{pattern:connector} that references an off-chain data source can be "tokenized" off-chain with access authorizations being encoded to a standard format using secure encryption and signing algorithms.  Types of algorithms employed along with public keys used to generate the tokens are captured by the attributes defined in the interface.  Tokens generated are then stored on-chain in a shared \textit{Token Registry} smart contract.  \textit{Token Registry} builds an audit trail of the creation, update, deletion, and access requests to each of the tokens.  To retrieve the \textit{Database Connector Object}, the recipient must possess the authorized party's secret key in order to decrypt and retrieve the original data source via the \textsc{Database Proxy} pattern presented in Section~\ref{pattern:proxy}.

\begin{figure}[tphb]
\begin{center}
\centerline{\includegraphics[width=0.7\columnwidth]{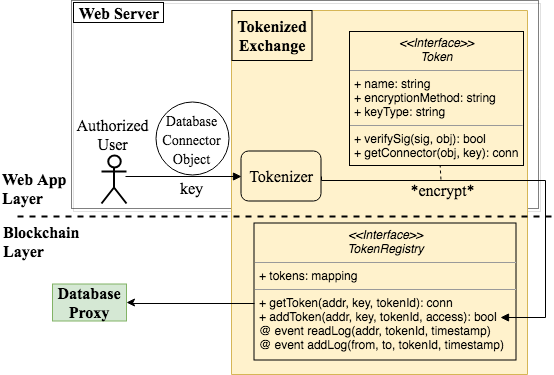}}
\caption{Structure of \textsc{Tokenized Exchange} Pattern for Authorizing off-Chain Data Access and Recording Verifiable Data Access Logs}
\label{fig:token}
\end{center}
\end{figure}

This approach ensure that even when tokens carrying actual information of a particular data source are shared with a wide network, they can only be consumed by the intended recipient(s) with proper cryptographically paired keys.  One drawback to this pattern is that there could be tokens that are not generalizable, in which case, implementations of other interfaces may be required.  Example interfaces include role-based access control models~\cite{sandhu1996role} and access matrix~\cite{sandhu1994access}, which provide more fine-grained authorizations and organizatin-specific rules to grant lower-level permissions to the access tokens.

\subsection{Providing Notifications of Relevant Healthcare Activities at Scale}
\label{pattern:pubsub}

\textbf{Design problem faced by blockchain-based apps.}  A blockchain-based healthcare system that needs to track relevant health changes across large patient populations must be designed to filter out useful health-related information from communication traffic (\textit{i.e.} transaction records) in the blockchain. For example, the Ethereum blockchain maintains an immutable record of contract creations and operation executions along with regular cryptocurrency transactions. The availability of this information makes blockchain a more autonomous approach to improve the coordination of patient care across different participants (\textit{e.g.}, physicians, pharmacists, insurance agents, etc) who would normally communicate through various channels with a lot of manual effort, such as through telephoning or faxing. Due to the continually growing list of records on the blockchain, however, directly capturing any specific health-related topic from occurred events implies exhaustive transaction receipt lookups and topic filtering, which requires non-trivial computation and may result in delayed responses. 

A good model should facilitate coordinated care and support relevant health information relays. For instance, health-related activities should be seamless communicated from the point when a patient self-reports illness (through a health DApp interface) to the point when they receive prescriptions created by their primary care provider; clinical reports and follow-up procedure should be relayed to and from the associated care provider offices in a timely manner. 

\textbf{Solution $\rightarrow$ Apply the \textit{Publisher-Subscriber} pattern to manage user notifications at scale when events of interest occur across the decentralized network.}
Incorporating a notification service using the \textit{Publisher-Subscriber} pattern~\cite{buschmann2007pattern} can facilitate scalable information filtering. In this design, changes in health activities are only broadcast to providers that subscribe to events relating to their patients. It alleviates tedious filtering of which care provider should be notified about patient activities as large volumes of transactions take place. It also helps maintain an interoperable environment that allows providers across various organizations to participate. 

Due to the deterministic nature of blockchain that supports smart contracts, communications between the on-chain address space and off-chain services can only occur in two ways.  The first way is a regular or constant poll, in which an off-chain server delegates a \textit{Messenger} component to monitor changes and new events in the system.  The second way pushes data out to an Oracle service, which is a trusted third-party that performs some computation off-chain and then forwards the results back to the blockchain address space via a callback function\footnote{https://blockchainhub.net/blockchain-oracles/}, such as in~\cite{oraclize2017}.  An Oracle service often charges a fee associated with its service provided to the blockchain and at its current stage today, it is not yet ideal for supporting large and sensitive data operations that are commonly experienced in a healthcare system.

The first variant avoids computation overhead on the blockchain because an off-chain server is responsible for querying and processing health events recorded on-chain.  Specifically, when the publisher sends an update, its subscribers only need to do a simple update to an internal state variable that records the publisher's address, which the DApp server delegates a \textit{Messenger} to actively monitor changes. When a change occurs, the responsibility for the heavy computational content filtering task (\textit{e.g.,} retrieving the change activity from the publisher using the address) is delegated to the DApp server from the blockchain. The DApp server is context-aware at this point because each subscriber has an associated contract address accessible by the server. The \textit{Messenger} can then filter the content based on subscribed topics and update the contract states of appropriate subscribers as needed.

The second variant shifts the responsibility of topic subscriptions and updates to the smart contract component on-chain.  When a topic, such as a patient their provider wishes to be notified of any health-related activities, experiences a new event or has a value update, the smart contract logic that notifies the subscribers pushes the updated topic to an Oracle service, which executes some task related to the topic (e.g., sending a secure message to the subscriber regarding the updated event) and sends the result back to the smart contract caller upon task completion.

\begin{figure}[th]
\begin{center}
\centerline{\includegraphics[width=0.7\columnwidth]{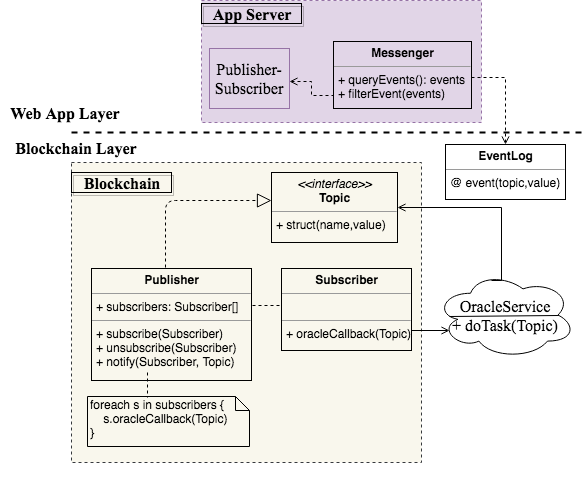}}
\caption{Two Variants of the \textsc{Publisher-Subscriber} Pattern for Providing Clinical Notifications of Relevant Healthcare Activities at Scale}
\label{fig:pubsub}
\end{center}
\end{figure}

Figure~\ref{fig:pubsub} shows the two variants of \textsc{Publisher-Subscriber} to provide the notification service. 

Implementing a notification service in a blockchain-based healthcare app is useful when a state change in the shared environment must be reported to interested parties without unmanaged many-to-many communications.  The disadvantage to the "poll" approach is the complexity in actual implementation of the messenger component that regularly monitors smart contract events, but it is much more efficient to unload the on-chain burden of topic filtering to off-chain services.  The drawbacks to the "push-to-oracle" approach are on-chain computation overhead and potential costs of Oracle services despite this approach being relatively easier to implement. 





\section{Conclusion}
\label{sec:conclusion}
Blockchain and programmable smart contracts provide an ecosystem for creating DApps that have the potential to improve healthcare interoperability on the technical level. However, due to the decentralization, immutability, and transparency properties of these technologies, a number of key concerns that especially arise in healthcare systems design must be addressed. These concerns include---but are not limited to---the evolvability of the system, balancing on-chain storage requirements and their overhead, the protection of patient data privacy, the scalability of the system across a large number of healthcare users, and information and system security protection. This chapter described these concerns and presented a key pattern sequence--\textsc{Layered Ring, Guarded Update, Contract Manager, Database Connector, Database Proxy, Entity Registry, Tokenized Exchange}, and \textsc{Publisher-Subscriber}--that together addresses these challenges.

Based on our experience developing the key pattern sequence, we learned the following lessons:  
\begin{itemize}
\item The public, immutable, and verifiable properties of the blockchain enable a more interoperable environment that is not easily achieved using traditional approaches that heavily rely on centralized systems. 
\item The modification of a smart contract is expensive both in terms of cost (if deployed on a public blockchain) and effort (that may require changes to be propagated to other system components). Important design decisions must therefore be made in advance to avoid the cost and storage overhead from changing the contract interface.
\item To best leverage key properties of blockchain and related technologies in the healthcare context, concerns regarding system evolvability, storage costs, sensitive information privacy, application scalability, and security protection must be taken into account.
\item Combining time-proven design practices with domain-knowledge that focus on better leveraging properties of blockchain technology enables the creation of systems that are more modular, easier to scale, less expensive to integrate and maintain, and less susceptible to change. 
\end{itemize}

\bibliographystyle{ACM-Reference-Format-Journals}
\bibliography{main}


\begin{thebibliography}{00}


\ifx \showCODEN    \undefined \def \showCODEN     #1{\unskip}     \fi
\ifx \showDOI      \undefined \def \showDOI       #1{{\tt DOI:}\penalty0{#1}\ }
  \fi
\ifx \showISBNx    \undefined \def \showISBNx     #1{\unskip}     \fi
\ifx \showISBNxiii \undefined \def \showISBNxiii  #1{\unskip}     \fi
\ifx \showISSN     \undefined \def \showISSN      #1{\unskip}     \fi
\ifx \showLCCN     \undefined \def \showLCCN      #1{\unskip}     \fi
\ifx \shownote     \undefined \def \shownote      #1{#1}          \fi
\ifx \showarticletitle \undefined \def \showarticletitle #1{#1}   \fi
\ifx \showURL      \undefined \def \showURL       #1{#1}          \fi

\bibitem[\protect\citeauthoryear{Ajami and Bagheri-Tadi}{Ajami and
  Bagheri-Tadi}{2013}]%
        {ajami2013barriers}
{Sima Ajami} {and} {Tayyebe Bagheri-Tadi}. 2013.
\newblock \showarticletitle{Barriers for adopting electronic health records
  (EHRs) by physicians}.
\newblock {\em Acta Informatica Medica\/} {21}, 2 (2013), 129.
\newblock


\bibitem[\protect\citeauthoryear{Atzei, Bartoletti, and Cimoli}{Atzei
  et~al\mbox{.}}{2017}]%
        {atzei2017survey}
{Nicola Atzei}, {Massimo Bartoletti}, {and} {Tiziana Cimoli}. 2017.
\newblock \showarticletitle{A survey of attacks on ethereum smart contracts
  (sok)}.
\newblock In {\em Principles of Security and Trust}. Springer, 164--186.
\newblock


\bibitem[\protect\citeauthoryear{Azaria, Ekblaw, Vieira, and Lippman}{Azaria
  et~al\mbox{.}}{2016}]%
        {azaria2016medrec}
{Asaph Azaria}, {Ariel Ekblaw}, {Thiago Vieira}, {and} {Andrew Lippman}. 2016.
\newblock \showarticletitle{Medrec: Using blockchain for medical data access
  and permission management}. In {\em Open and Big Data (OBD), International
  Conference on}. IEEE, 25--30.
\newblock


\bibitem[\protect\citeauthoryear{Bartoletti and Pompianu}{Bartoletti and
  Pompianu}{2017}]%
        {bartoletti2017empirical}
{Massimo Bartoletti} {and} {Livio Pompianu}. 2017.
\newblock \showarticletitle{An empirical analysis of smart contracts:
  platforms, applications, and design patterns}.
\newblock {\em arXiv preprint arXiv:1703.06322\/} (2017).
\newblock


\bibitem[\protect\citeauthoryear{Blundell-Wignall}{Blundell-Wignall}{2014}]%
        {blundell2014bitcoin}
{Adrian Blundell-Wignall}. 2014.
\newblock \showarticletitle{The Bitcoin question: Currency versus trust-less
  transfer technology}.
\newblock {\em OECD Working Papers on Finance, Insurance and Private
  Pensions\/} 37 (2014), 1.
\newblock


\bibitem[\protect\citeauthoryear{Broderson, Kalis, Leong, Mitchell, Pupo, and
  Truscott}{Broderson et~al\mbox{.}}{2016}]%
        {Broderson2016}
{C Broderson}, {B Kalis}, {C Leong}, {E Mitchell}, {E Pupo}, {and} {A
  Truscott}. 2016.
\newblock Blockchain: Securing a New Health Interoperability Experience.
\newblock   (2016).
\newblock


\bibitem[\protect\citeauthoryear{Buschmann, Henney, and Schimdt}{Buschmann
  et~al\mbox{.}}{2007}]%
        {buschmann2007pattern}
{Frank Buschmann}, {Kelvin Henney}, {and} {Douglas Schimdt}. 2007.
\newblock {\em Pattern-oriented Software Architecture: on patterns and pattern
  language}. Vol.~5.
\newblock John wiley \& sons.
\newblock


\bibitem[\protect\citeauthoryear{Buterin et~al\mbox{.}}{Buterin
  et~al\mbox{.}}{2013}]%
        {buterin2013ethereum}
{Vitalik Buterin} {and} {others}. 2013.
\newblock Ethereum white paper.
\newblock   (2013).
\newblock


\bibitem[\protect\citeauthoryear{Cap}{Cap}{nd}]%
        {cap2018cryptocurrency}
{Coin~Market Cap}. n.d.
\newblock Cryptocurrency Market Capitalizations.
\newblock   (n.d.).
\newblock


\bibitem[\protect\citeauthoryear{CDC}{CDC}{2003}]%
        {centers2003hipaa}
{CDC}. 2003.
\newblock \showarticletitle{HIPAA privacy rule and public health. Guidance from
  CDC and the US Department of Health and Human Services}.
\newblock {\em MMWR: Morbidity and mortality weekly report\/} {52}, Suppl. 1
  (2003), 1--17.
\newblock


\bibitem[\protect\citeauthoryear{ConsenSys}{ConsenSys}{2018}]%
        {consensys2018practice}
{ConsenSys}. 2018.
\newblock \showarticletitle{Recommendations for Smart Contract Security in
  Solidity}.
\newblock {\em Web. Recommendations for Smart Contract Security in Solidity\/}
  (2018).
\newblock


\bibitem[\protect\citeauthoryear{Coplien, Hoffman, and Weiss}{Coplien
  et~al\mbox{.}}{1998}]%
        {coplien1998commonality}
{James Coplien}, {Daniel Hoffman}, {and} {David Weiss}. 1998.
\newblock \showarticletitle{Commonality and variability in software
  engineering}.
\newblock {\em IEEE software\/} {15}, 6 (1998), 37--45.
\newblock


\bibitem[\protect\citeauthoryear{CryptoKitties}{CryptoKitties}{nd}]%
        {cryptokitties}
CryptoKitties n.d.
\newblock CryptoKitties.
\newblock   (n.d.).
\newblock


\bibitem[\protect\citeauthoryear{Das}{Das}{2017}]%
        {das2017does}
{Reenita Das}. 2017.
\newblock Does Blockchain Have A Place In Healthcare.
\newblock   (2017).
\newblock


\bibitem[\protect\citeauthoryear{DeSalvo and Galvez}{DeSalvo and
  Galvez}{2015}]%
        {desalvo2015connecting}
{K DeSalvo} {and} {E Galvez}. 2015.
\newblock \showarticletitle{Connecting health and care for the nation: a shared
  nationwide interoperability roadmap—version 1.0}.
\newblock {\em Health IT Buzz\/} (2015).
\newblock


\bibitem[\protect\citeauthoryear{Dourlens}{Dourlens}{2017}]%
        {dourlens}
{Jules Dourlens}. 2017.
\newblock Ethereum smart contracts lifecycle.
\newblock   (2017).
\newblock


\bibitem[\protect\citeauthoryear{Dubovitskaya, Xu, Ryu, Schumacher, and
  Wang}{Dubovitskaya et~al\mbox{.}}{2017}]%
        {Dubovitskaya2017}
{Alevtina Dubovitskaya}, {Zhigang Xu}, {Samuel Ryu}, {Michael Schumacher},
  {and} {Fusheng Wang}. 2017.
\newblock \showarticletitle{Secure and Trustable Electronic Medical Records
  Sharing using Blockchain}.
\newblock {\em arXiv preprint arXiv:1709.06528\/} (2017).
\newblock


\bibitem[\protect\citeauthoryear{Ellervee, Matulevicius, and Mayer}{Ellervee
  et~al\mbox{.}}{2017}]%
        {ellervee2017comprehensive}
{Andreas Ellervee}, {Raimundas Matulevicius}, {and} {Nicolas Mayer}. 2017.
\newblock \showarticletitle{A Comprehensive Reference Model for
  Blockchain-based Distributed Ledger Technology.}. In {\em ER Forum/Demos}.
  306--319.
\newblock


\bibitem[\protect\citeauthoryear{Etherescan}{Etherescan}{nd}]%
        {ethscan}
Etherescan n.d.
\newblock Etherescan - The Ethereum Blockchain Explorer.
\newblock   (n.d.).
\newblock


\bibitem[\protect\citeauthoryear{Ethereum.io}{Ethereum.io}{2017}]%
        {soliditypoly}
{Ethereum.io}. 2017.
\newblock \showarticletitle{Contracts}.
\newblock {\em Web.
  \url{http://solidity.readthedocs.io/en/develop/contracts.html}\/} (2017).
\newblock


\bibitem[\protect\citeauthoryear{Fernandez}{Fernandez}{2013}]%
        {fernandez2013security}
{Eduardo~B. Fernandez}. 2013.
\newblock {\em Security patterns in practice: designing secure architectures
  using software patterns}.
\newblock John Wiley \& Sons.
\newblock


\bibitem[\protect\citeauthoryear{FOMO3D}{FOMO3D}{nd}]%
        {fomo3d}
FOMO3D n.d.
\newblock FOMO3D.
\newblock   (n.d.).
\newblock


\bibitem[\protect\citeauthoryear{Foundation}{Foundation}{2015a}]%
        {oraclize2017}
{Ethereum Foundation}. 2015a.
\newblock \showarticletitle{ORACLIZE LIMITED}.
\newblock {\em Web. \url{http://www.oraclize.it/}\/} (2015).
\newblock


\bibitem[\protect\citeauthoryear{Foundation}{Foundation}{2015b}]%
        {solidity2017ethereum}
{Ethereum Foundation}. 2015b.
\newblock \showarticletitle{Solidity}.
\newblock {\em Web. \url{https://solidity.readthedocs.io/en/develop/}\/}
  (2015).
\newblock


\bibitem[\protect\citeauthoryear{Gamma, Helm, Johnson, and Vlissides}{Gamma
  et~al\mbox{.}}{1993}]%
        {gamma1993design}
{Erich Gamma}, {Richard Helm}, {Ralph Johnson}, {and} {John Vlissides}. 1993.
\newblock \showarticletitle{Design patterns: Abstraction and reuse of
  object-oriented design}. In {\em European Conference on Object-Oriented
  Programming}. Springer, 406--431.
\newblock


\bibitem[\protect\citeauthoryear{Gamma, Vlissides, Johnson, and Richard}{Gamma
  et~al\mbox{.}}{1995}]%
        {gamma1995design}
{Erich Gamma}, {John Vlissides}, {Ralph Johnson}, {and} {Helm. Richard}. 1995.
\newblock {\em Design Patterns: Elements of Reusable Object-Oriented Software}.
\newblock Pearson Education.
\newblock


\bibitem[\protect\citeauthoryear{Geraci, Katki, McMonegal, Meyer, Lane, Wilson,
  Radatz, Yee, Porteous, and Springsteel}{Geraci et~al\mbox{.}}{1991}]%
        {geraci1991ieee}
{Anne Geraci}, {Freny Katki}, {Louise McMonegal}, {Bennett Meyer}, {John Lane},
  {Paul Wilson}, {Jane Radatz}, {Mary Yee}, {Hugh Porteous}, {and} {Fredrick
  Springsteel}. 1991.
\newblock {\em IEEE standard computer dictionary: Compilation of IEEE standard
  computer glossaries}.
\newblock IEEE Press.
\newblock


\bibitem[\protect\citeauthoryear{Hub}{Hub}{2017}]%
        {cmublockchain}
{Blockchain Hub}. 2017.
\newblock Blockchain Oracles.
\newblock Web.
  \url{https://insights.sei.cmu.edu/sei_blog/2017/07/what-is-bitcoin-what-is-blockchain.html}.
    (2017).
\newblock


\bibitem[\protect\citeauthoryear{IDEX - Decentralized Ethereum Asset
  Exchange}{IDEX - Decentralized Ethereum Asset Exchange}{2018}]%
        {idex}
IDEX - Decentralized Ethereum Asset Exchange 2018.
\newblock IDEX - Decentralized Ethereum Asset Exchange.
\newblock   (2018).
\newblock


\bibitem[\protect\citeauthoryear{Johnston, Yilmaz, Kandah, Bentenitis, Hashemi,
  Gross, Wilkinson, and Mason}{Johnston et~al\mbox{.}}{2014}]%
        {johnston2014general}
{David Johnston}, {Sam~Onat Yilmaz}, {Jeremy Kandah}, {Nikos Bentenitis},
  {Farzad Hashemi}, {Ron Gross}, {Shawn Wilkinson}, {and} {Steven Mason}. 2014.
\newblock \showarticletitle{The General Theory of Decentralized Applications,
  DApps}.
\newblock {\em GitHub, June\/}  {9} (2014).
\newblock


\bibitem[\protect\citeauthoryear{Lesh, Weininger, Goldman, Wilson, and
  Himes}{Lesh et~al\mbox{.}}{2007}]%
        {lesh2007medical}
{Kathy Lesh}, {Sandy Weininger}, {Julian~M Goldman}, {Bob Wilson}, {and} {Glenn
  Himes}. 2007.
\newblock \showarticletitle{Medical device interoperability-assessing the
  environment}. In {\em 2007 Joint Workshop on High Confidence Medical Devices,
  Software, and Systems and Medical Device Plug-and-Play Interoperability
  (HCMDSS-MDPnP 2007)}. IEEE, 3--12.
\newblock


\bibitem[\protect\citeauthoryear{Moreno, Fernandez, Fernandez-Medina, and
  Serrano}{Moreno et~al\mbox{.}}{2019}]%
        {moreno2019}
{Julio Moreno}, {Eduardo~B. Fernandez}, {Eduardo Fernandez-Medina}, {and}
  {Manuel Serrano}. 2019.
\newblock \showarticletitle{A Security Pattern to Incorporate Blockchain in Big
  Data Ecosystems}. In {\em EuroPLoP-24th European Conference on Pattern
  Languages of Programs}.
\newblock


\bibitem[\protect\citeauthoryear{Nakamoto}{Nakamoto}{2008}]%
        {nakamoto2008bitcoin}
{Satoshi Nakamoto}. 2008.
\newblock Bitcoin: A peer-to-peer electronic cash system.
\newblock   (2008).
\newblock


\bibitem[\protect\citeauthoryear{ONC}{ONC}{2014}]%
        {onc2014vision}
{ONC}. 2014.
\newblock Connecting Health and Care for the Nation: A 10-Year Vision to
  Achieve an Interoperable Health IT Infrastructure.
\newblock   (2014).
\newblock


\bibitem[\protect\citeauthoryear{Palladino}{Palladino}{2017}]%
        {zepplin2018parity}
{Santiago Palladino}. 2017.
\newblock \showarticletitle{The Parity Wallet Hack Explained}.
\newblock {\em Web.
  \url{https://blog.zeppelin.solutions/on-the-parity-wallet-multisig-hack-405a8c12e8f7}\/}
  (2017).
\newblock


\bibitem[\protect\citeauthoryear{Peter B.~Nichol}{Peter B.~Nichol}{2016}]%
        {Nichol2016}
{Jeff~Brandt Peter B.~Nichol}. 2016.
\newblock Co-Creation of Trust for Healthcare: The Cryptocitizen. Framework for
  Interoperability with Blockchain.
\newblock   (2016).
\newblock


\bibitem[\protect\citeauthoryear{Peterson, Deeduvanu, Kanjamala, and
  Boles}{Peterson et~al\mbox{.}}{2016}]%
        {Peterson2016}
{Kevin Peterson}, {Rammohan Deeduvanu}, {Pradip Kanjamala}, {and} {Kelly
  Boles}. 2016.
\newblock A Blockchain-Based Approach to Health Information Exchange Networks.
\newblock   (2016).
\newblock


\bibitem[\protect\citeauthoryear{Porru, Pinna, Marchesi, and Tonelli}{Porru
  et~al\mbox{.}}{2017}]%
        {porru2017blockchain}
{Simone Porru}, {Andrea Pinna}, {Michele Marchesi}, {and} {Roberto Tonelli}.
  2017.
\newblock \showarticletitle{Blockchain-oriented software engineering:
  challenges and new directions}. In {\em Proceedings of the 39th International
  Conference on Software Engineering Companion}. IEEE Press, 169--171.
\newblock


\bibitem[\protect\citeauthoryear{Rich and Gellman}{Rich and Gellman}{2014}]%
        {rich2014nsa}
{Steven Rich} {and} {Barton Gellman}. 2014.
\newblock \showarticletitle{NSA seeks to build quantum computer that could
  crack most types of encryption}.
\newblock {\em The Washington Post\/}  {2} (2014).
\newblock


\bibitem[\protect\citeauthoryear{Richesson and Nadkarni}{Richesson and
  Nadkarni}{2011}]%
        {richesson2011data}
{Rachel~L Richesson} {and} {Prakash Nadkarni}. 2011.
\newblock \showarticletitle{Data standards for clinical research data
  collection forms: current status and challenges}.
\newblock {\em Journal of the American Medical Informatics Association\/} {18},
  3 (2011), 341--346.
\newblock


\bibitem[\protect\citeauthoryear{Sandhu, Coyne, Feinstein, and Youman}{Sandhu
  et~al\mbox{.}}{1996}]%
        {sandhu1996role}
{Ravi~S Sandhu}, {Edward~J Coyne}, {Hal~L Feinstein}, {and} {Charles~E Youman}.
  1996.
\newblock \showarticletitle{Role-based access control models}.
\newblock {\em Computer\/} {29}, 2 (1996), 38--47.
\newblock


\bibitem[\protect\citeauthoryear{Sandhu and Samarati}{Sandhu and
  Samarati}{1994}]%
        {sandhu1994access}
{Ravi~S Sandhu} {and} {Pierangela Samarati}. 1994.
\newblock \showarticletitle{Access control: principle and practice}.
\newblock {\em IEEE communications magazine\/} {32}, 9 (1994), 40--48.
\newblock


\bibitem[\protect\citeauthoryear{Shvets}{Shvets}{2015}]%
        {shvets2015design}
{Alexander Shvets}. 2015.
\newblock \showarticletitle{Design Patterns Explained Simply}.
\newblock {\em sourcemaking. com\/} (2015), 80--84.
\newblock


\bibitem[\protect\citeauthoryear{Siegel}{Siegel}{2016}]%
        {siegel2016understanding}
{David Siegel}. 2016.
\newblock \showarticletitle{Understanding the DAO attack}.
\newblock {\em Web.
  \url{http://www.coindesk.com/understanding-dao-hack-journalists}\/} (2016).
\newblock


\bibitem[\protect\citeauthoryear{Xu, Pautasso, Zhu, Gramoli, Ponomarev, Tran,
  and Chen}{Xu et~al\mbox{.}}{2016}]%
        {xu2016blockchain}
{Xiwei Xu}, {Cesare Pautasso}, {Liming Zhu}, {Vincent Gramoli}, {Alexander
  Ponomarev}, {An~Binh Tran}, {and} {Shiping Chen}. 2016.
\newblock \showarticletitle{The blockchain as a software connector}. In {\em
  2016 13th Working IEEE/IFIP Conference on Software Architecture (WICSA)}.
  IEEE, 182--191.
\newblock


\bibitem[\protect\citeauthoryear{Zdun, Hentrich, and Van Der~Aalst}{Zdun
  et~al\mbox{.}}{2006}]%
        {zdun2006survey}
{Uwe Zdun}, {Carsten Hentrich}, {and} {Wil~MP Van Der~Aalst}. 2006.
\newblock \showarticletitle{A survey of patterns for service-oriented
  architectures}.
\newblock {\em International journal of Internet protocol technology\/} {1}, 3
  (2006), 132--143.
\newblock


\bibitem[\protect\citeauthoryear{Zhang, Schmidt, White, and Lenz}{Zhang
  et~al\mbox{.}}{2018}]%
        {ZHANG2018}
{Peng Zhang}, {Douglas~C. Schmidt}, {Jules White}, {and} {Gunther Lenz}. 2018.
\newblock \showarticletitle{Blockchain Technology Use Cases in Healthcare}.
\newblock In {\em Blockchain Technology: Platforms, Tools, and Use Cases}.
  Elsevier.
\newblock
\showISSN{0065-2458}
\showDOI{%
\url{http://dx.doi.org/https://doi.org/10.1016/bs.adcom.2018.03.006}}


\bibitem[\protect\citeauthoryear{Zhang, Walker, White, Schmidt, and Lenz}{Zhang
  et~al\mbox{.}}{2017a}]%
        {zhang2017metric}
{Peng Zhang}, {Michael~A Walker}, {Jules White}, {Douglas~C Schmidt}, {and}
  {Gunther Lenz}. 2017a.
\newblock \showarticletitle{Metrics for assessing blockchain-based healthcare
  decentralized apps}. In {\em 2017 IEEE 19th International Conference on
  e-Health Networking, Applications and Services (Healthcom)}. IEEE, 1--4.
\newblock


\bibitem[\protect\citeauthoryear{Zhang, White, Schmidt, and Lenz}{Zhang
  et~al\mbox{.}}{2017b}]%
        {zhang2017applying}
{Peng Zhang}, {Jules White}, {Douglas~C Schmidt}, {and} {Gunther Lenz}. 2017b.
\newblock \showarticletitle{Design of blockchain-based apps using familiar
  software patterns with a healthcare focus}. In {\em Proceedings of the 24th
  Conference on Pattern Languages of Programs}. The Hillside Group, 19.
\newblock


\bibitem[\protect\citeauthoryear{Zhang, White, Schmidt, Lenz, and
  Rosenbloom}{Zhang et~al\mbox{.}}{2018}]%
        {ZHANG2018267}
{Peng Zhang}, {Jules White}, {Douglas~C Schmidt}, {Gunther Lenz}, {and}
  {S~Trent Rosenbloom}. 2018.
\newblock \showarticletitle{FHIRChain: applying blockchain to securely and
  scalably share clinical data}.
\newblock {\em Computational and structural biotechnology journal\/}  {16}
  (2018), 267--278.
\newblock


\end{thebibliography}

\received{June 2019}{September 2019}{February 2020}

\end{document}